%% file: main.tex
\documentclass[conference]{IEEEtran}
\IEEEoverridecommandlockouts
\usepackage{comment}
\usepackage{cite}
\usepackage{amsmath,amssymb,amsfonts}
\usepackage{algorithmic}
\usepackage{graphicx}
\usepackage{textcomp}
\usepackage{xcolor}
\usepackage{caption}
\usepackage{subcaption}
\usepackage[normalem]{ulem}
\usepackage{soul}
\usepackage{booktabs}
\usepackage{multicol}
\usepackage{amsfonts}

\def\BibTeX{{\rm B\kern-.05em{\sc i\kern-.025em b}\kern-.08em
    T\kern-.1667em\lower.7ex\hbox{E}\kern-.125emX}}
    
\usepackage{fancyhdr}
\usepackage{kantlipsum}

\fancypagestyle{plain}{
\fancyhf{}
\fancyhead[C]{Conference on \LaTeX} 

}
\usepackage{eso-pic}
\begin{document}

\AddToShipoutPictureBG*{
\AtPageUpperLeft{

\setlength\unitlength{1in}

\hspace*{\dimexpr0.5\paperwidth\relax}

\makebox(0,-0.75)[c]{\textbf{2021 IEEE/ACM International Conference on Advances in Social Networks Analysis and Mining (ASONAM)}}
}}

\title{Overall Behavioural Index (OBI) For Measuring Segregation}

\author{\IEEEauthorblockN{Rahul Goel}
\IEEEauthorblockA{
\textit{Institute of Computer Science}\\
University of Tartu, Estonia \\
rahul.goel@ut.ee}
\and
\IEEEauthorblockN{Rajesh Sharma}
\IEEEauthorblockA{\textit{Institute of Computer Science} \\
University of Tartu, Estonia\\
rajesh.sharma@ut.ee}
\and
\IEEEauthorblockN{Anto Aasa}
\IEEEauthorblockA{\textit{ Department of Geography} \\
University of Tartu, Estonia\\
anto.aasa@ut.ee}
}

\maketitle
\IEEEoverridecommandlockouts
\IEEEpubid{\parbox{\columnwidth}{\vspace{8pt}
\makebox[\columnwidth][t]{IEEE/ACM ASONAM 2021, November 8-11, 2021}
\makebox[\columnwidth][t]{978-1-????-????-1/20/\$31.00~\copyright\space2021 IEEE} \hfill} \hspace{\columnsep}\makebox[\columnwidth]{}}
\IEEEpubidadjcol
\begin{abstract}  
Segregation, defined as the degree of separation between two or more population groups, helps to understand a complex social environment and subsequently provides a basis for public policy intervention. To measure segregation, past works often propose indexes that are criticized for being over-simplified and over-reduced. In other words, these indexes use the highly aggregated information to measure segregation. In this paper, we propose three novel indexes to measure segregation, namely: (i) \textit{Individual Segregation Index (ISI)}, (ii) \textit{Individual Inclination Index (III)}, and (iii) \textit{Overall Behavioural Index (OBI)}. The \textit{ISI} index measures individuals' segregation, and the \textit{III} index reports the individuals' inclination towards other population groups. The \textit{OBI} index, calculated using both \textit{III} and \textit{ISI} index, is non-simplified and not only recognizes individuals' connectivity behaviour but group's connectivity behavioural distribution as well. By considering commonly used \textit{Freeman's segregation} and \textit{homophily index} as baseline indexes, we compare the \textit{OBI} index on real call data records (CDR) dataset of Estonia to show the effectiveness of the proposed indexes.

\end{abstract}


\input{Intro}

\input{related}
\input{Dataset}
\input{Methodology}

\input{Eval}

\input{Conclusion}

\section*{Acknowledgment}
This research is funded by ERDF via the IT Academy Research Programme and H2020 Project, SoBigData++ and CHIST-ERA project SAI.
\vspace{-3mm}
\bibliographystyle{IEEEtran}
\bibliography{Seeders}

\end{document}

%% file: Intro.tex
\section{Introduction}
\label{sec:intro}

Over many decades, studying segregation in society has attracted a lot of attention from the research community \cite{farley1994changes,goel2021understanding}. At its core, studying segregation can reveal information about weak communities in a society \cite{oliver1999effects} and identifying them is essential as they are socially and politically vulnerable \cite{meister2020comparing}. Also, communities that are segregated often experience linguistic isolation \cite{santiago2019framework}, which makes it more difficult for them to obtain jobs, education, health care, and integrate with other communities 
\cite{ward2018neighborhood}. Thus, segregation study plays a crucial role by summarizing a complex social environment and providing ground for public policy intervention. Studying segregation between populations of different ages, ethnicities, and social classes have been a prominent theme for at least half a century 
\cite{goedel2020association}. The most well-known index to measure segregation, the \textit{dissimilarity index}, came to popularity following the work of \cite{duncan1955methodological} which demonstrated the strength of the index compared to alternate measures available at that time \cite{massey1988dimensions}. 
Later, following the index's criticism in \cite{cortese1976further}, a vast number of works start offering alternative methods and indexes for measuring the degree of separation between two or more population groups \cite{massey1988dimensions}. However, these indexes are often criticized for being over-simplified and over-reduced. That means to measure segregation these indexes use highly aggregated information, such as the number of individuals in each group and the ties within and in-between groups.

To overcome the limitations of the previous works, we propose the following three novel indexes to measure segregation:

\begin{enumerate}
    \item \textbf{Individual Segregation Index (ISI)}: The \textit{ISI} index measures the segregation for each individual considering the whole population distribution. For example, if an individual only connects to minority group individuals, then according to \textit{ISI} index, s/he is more segregated than an individual only connected to majority group individuals. However, this index is not adequate to explain an individual's actual behaviour because if an individual is segregated, then the next relevant question is to which group s/he prefers to connect (or 
    in other words is more inclined). 
    We measure this inclination using our second proposed index, the \textit{III} index.

    \item \textbf{Individual Inclination Index (III)}: As its name implies, the \textit{III} index measures an individual's connection preferences, keeping in mind the whole population distribution. E.g., if an individual prefers to connect within the same group, their \textit{III} index value will be positive. On the other hand, if an individual prefers to connect with other group(s), their \textit{III} index will be negative.

    \item \textbf{Overall Behavioural Index (OBI)}: The \textit{OBI} index is determined using both the \textit{ISI} and the \textit{III} indexes and identify the behaviour of individuals. The \textit{OBI} index considers the segregation and inclination of each individual in the population; thus, it is more reliable than other existing indexes.
\end{enumerate}



We investigated the past segregation indexes and find out that they are inconsistent with each other, in the sense that the relative order of segregation is not preserved, e.g., according to index $I_1$, group $G_1$ is more segregated than $G_2$, but according to index $I_2$, $G_2$ is more segregated than $G_1$. Another drawback of these indexes is that they can not explain the individuals' connectivity behaviour that results in segregation, which we overcome by using the \textit{OBI} index. In case, the \textit{OBI} index returns a segregation value, it is possible to examine what connectivity behaviour of individuals leads to this segregation with the help of the group's behavioural distribution, which is the contribution of our work.

%% file: related.tex
\section{Related work}
In this section, we discuss relevant literature with respect to different segregation measures which involves two different lines of work. First involving non-spatial indices, in which the information is independent of all geographic considerations 
(Section \ref{subsec:non-spatial}). Second set is referred to as spatial indices which are defined as those which directly or indirectly consider location information \cite{wong1993spatial} (Section \ref{subsec:spatial}).

\begin{table}
\centering
\begin{tabular}{|l|p{55mm}|l|} 
\hline
\textbf{S.No.} & \textbf{Index}   & \textbf{Citation}   \\ 
\hline
\multicolumn{3}{|c|}{\textbf{\textcolor{blue}{Non-Spatial Indices}}} \\ 
\hline
1 & Dissimilarity index           & \cite{duncan1955methodological}           \\ 
\hline
2 & Gini index       & \cite{duncan1955methodological}          \\ 
\hline
3 & Centralisation index          & \cite{duncan1955residential}           \\ 
\hline
4 & Coleman’s Homophily index     & \cite{coleman1958relational}       \\ 
\hline
5 & Freeman’s index  & \cite{freeman1978segregation}       \\ 
\hline
6 & Multi–group dissimilarity index            & \cite{morgan1975segregation} \\  
\hline
7 & Exposure index   & \cite{lieberson1981asymmetrical}          \\ 
\hline
8 & Neighbourhood sorting index (NSI)          & \cite{jargowsky1996take}            \\ 
\hline
9 & Typology for classifying ethnic residential areas       & \cite{poulsen2001intraurban} \\ 
\hline
10& Location Quotient (LQ)        & \cite{brown2006spatial}         \\ 
\hline
\multicolumn{3}{|c|}{\textbf{\textcolor{blue}{Spatial Indices}}}     \\ 
\hline
11& Spatial proximity (SP) index  &      \cite{white1983measurement}  \\ 
\hline
12& Multi-group SP   &   \cite{grannis2002discussion}     \\ 
\hline
13& Dissimilarity index incorporating spatial adjacency     &  \cite{morrill1991measure}      \\ 
\hline
14& Dssimilarity index incorporating common boundary lengths&   \cite{wong1993spatial}     \\ 
\hline
15& Dssimilarity index incorporating common boundary lengths and perimeter/area ratio &  \cite{reardon2004measures}     \\
\hline
16& Spatial version of multigroup dissimilarity index       &    \cite{horn2005measuring}    \\ 
\hline
17& General index of spatial segregation       &    \cite{wong2005formulating}    \\ 
\hline
18& Generalised spatial dissimilarity (GSD) index           &     \cite{feitosa2007global}   \\ 
\hline
19& Spatial dissimilarity index   &   \cite{reardon2004measures}       \\ 
\hline
20& Spatial information theory index           &   \cite{reardon2004measures}       \\ 
\hline
21& Local GSD        &    \cite{feitosa2007global}      \\ 
\hline
22& focal LQ         &   \cite{cromley2012focal}     \\ 
\hline
23& Local Moran’s I  &    \cite{anselin1995local}    \\ 
\hline
24& Getis–Ord local G& \cite{getis2010analysis}       \\
\hline
\end{tabular}
\caption{Non-spatial and spatial segregation indices.}
\label{table:Indices}
\end{table}

\subsection{Non-Spatial Indices}\label{subsec:non-spatial}
The one most well-known non-spatial index 
is the dissimilarity index D \cite{duncan1955methodological}.

\begin{equation}\label{eq:D1}
    D = \frac{1}{2} \sum_i \left| \frac{p_{i,g}}{p_g}  - \frac{p_{i,{\bar{g}}}}{p_{\bar{g}}} \right|
\end{equation}


where $i$ is the index of spatial unit; g, $\bar{\mbox{g}}$ represent two population groups; $p_g$, $p_{\bar{g}}$ are total population of the two groups in the entire study region; $p_{i,g}$, $p_{i,\bar{g}}$ are population of groups g, $\bar{\mbox{g}}$ in spatial unit $i$, respectively. 

There exists a few similar measures like D 
summarized in Table \ref{table:Indices} under non-spatial indices (S.No. 2-10). All these measures share several significant limitations. First, they cast out a great deal of geographical detail, considering the residential system as consisting of different entities, each area being isolated from neighboring areas. Second, they focus on a global summary for a city or region, assuming that spatial relations are consistent across that area. Third, they look at the residential space without paying attention to the varied locations in which people spend time during the day.

\subsection{Spatial Indices}\label{subsec:spatial}
There are several extensions of D that incorporate spatial relationships between various groups of population in measuring segregation index. For example, Spatial proximity (SP) index \cite{white1983measurement}, Multi-group SP \cite{grannis2002discussion}, etc (see Table \ref{table:Indices}, S.No. 11-24).

In these indices, geographical distance between the population groups is a fundamental metric to understand the spatial relationship between them. For example, in \cite{white1983measurement}, authors utilize a distance function to measure social interaction changes with distance. This distance metric can be reduced to an adjacency matrix, which includes spatial connectivity information. In a different work \cite{morrill1991measure}, the author introduced a spatial adjacency term $w_{ij}$ in $D$ that represents the spatial connections among various population groups. This adjacency index is represented as $D(adj)$ and can be defined as follows:
\begin{equation}
    D(adj) = D-\frac{\sum_i \sum_j |w_{ij}(x_{i,g}-x_{j,g})|}{\sum_i \sum_j w_{ij}}
\end{equation}

where, $x_{i,g}$ and $x_{j,g}$ represents the proportion of population of group $g$ in spatial units $i$ and $j$ respectively. If $i$ and $j$ are adjacent to each other, the value of $w_{ij}$ 
is 1 otherwise 0. 

Later, in \cite{wong1993spatial} author suggested two updated versions of D(adj): D(w), which assumes that longer mutual boundaries between units allow more spatial interaction, and D(s), which assumes that the compactness of neighboring units influences interaction between units. However, there is a functional drawback to the Dissimilarity Index and its spatial versions 
as they compare only two groups, whereas many populations 
can have several 
groups. 
A variant for multiple groups, D(m), was suggested in \cite{morgan1975segregation}. 
Similar to D, D(m) implies no contact in neighboring units between populations and is therefore aspatial. The spatial variant, SD(m) was suggested in \cite{wong1998measuring}, which was structurally similar to D(m).

In this work, we propose a non-spatial segregation index \textbf{O}verall \textbf{B}ehavioural \textbf{I}ndex (OBI) which is a combination of \textit{Individual Segregation Index (ISI)} and \textit{Individual Inclination Index (III)} that reports the segregation as well as inclination of the population group under investigation.

%% file: Dataset.tex
\section{Dataset Description}\label{sec:ds}
This section takes a closer look at the dataset that we will use in Section \ref{sec:methodology} for explaining our proposed indexes and in Section \ref{sec:Evl} for comparing our proposed segregation index with baseline indexes. This study utilizes anonymized call data records (\textit{CDR}) issued by one of Estonia's leading mobile operators. The dataset contains timestamp data to the level of seconds for each call operation and the cell phone tower's passive mobile location. The call records span six days, from $8^{th}$ May 2017 to $13^{th}$ May 2017. The data collection consists of 12,317,970 independent call records from 1,175,191 unique individuals, which is 89.32\% population of Estonia \cite{estonia2012statistical}.

The dataset contains following information for each call activity: 
user \textit{pseudonymous ID}, \textit{timestamp} (with a precision of 1 second), and \textit{location} of the network cell. The pseudonymous ID ensures the user's anonymity. 
In addition, the \textit{gender} of the user, \textit{year of birth} and preferred language of communication are provided in the dataset for research purposes. The preferred language of interaction choices is either \textit{Estonian}, \textit{Russian}, or \textit{English}, as selected by the customer when signing the contract with the service provider. It is to be noted that not all users have additional details (gender, language, and location) in the dataset. For example, \textit{gender} information is available for 130,988 users with 61,933 males and 69,055 females. Table \ref{Table:DataSetStats} summarizes statistics for this dataset. 

\begin{table}
\centering
\begin{tabular}{lcccc}
\toprule
    \textbf{Parameters}  & \textbf{Value}& \textbf{FSI} & \textbf{HI} & \textbf{OBI}\\ \midrule
 Time period & 8 May'17 to 13 May'17 \\ \midrule
 Call Records & 12,179,970 \\ \midrule
 Unique Users & 1,175,919 \\ \midrule
    \multicolumn{5}{@{}l}{\textbf{Gender}}\\
    Male & 61,933  & 0.267  & 0.158 & 0.153 \\ 
    Female & 69,055 &    0.267  & 0.1 & 0.087  \\\midrule
     \multicolumn{2}{@{}l}{\textbf{Age-Groups}}\\
    (0,14]  & 76            &   0.374    & 0.032 & -0.727 \\ 
    (14,24] & 1,196         &   0.206   & 0.172 & -0.340 \\
    (24,54] & 83,028        &   0.385    & \textbf{0.561} & \textbf{0.552} \\
    (54,64] & 21,427        &   0.322   & 0.398 & 0.314 \\
    (64,100] & 12,323        &  \textbf{0.432}    & 0.332 & 0.079 \\
  \midrule
   \multicolumn{5}{@{}l}{\textbf{Languages}}\\
    Estonian &  102,545     &  0.738    & 0.704 & 0.488 \\
    Russian &  14,882       &   \textbf{0.742}   & \textbf{0.751} & \textbf{0.527} \\
    English &  236          &  0.267  & 0.094 & -0.463 \\
  \midrule
   \multicolumn{5}{@{}l}{\textbf{Locations}}\\
   Harju & 507,365      &   0.712 & 0.798 & 0.487 \\
    Hiiu & 6,959        &   0.812  & 0.724 & 0.574 \\
    Ida-Viru & 87,212   &   \textbf{0.856}   & \textbf{0.838} & \textbf{0.754} \\
    Järva &  18,157     &   0.762   & 0.537 & 0.448 \\
    Jõgeva &  23,125    &   0.758  & 0.587 & 0.542 \\
    Lääne &  18,377     &   0.773   & 0.608 & 0.514 \\
    Lääne-Viru & 44,787 &   0.778    & 0.695 & 0.627 \\
    Pärnu &  55,873     &   0.796   & 0.705 & 0.645 \\
    Põlva & 21,083      &   0.754    & 0.591 & 0.532 \\
    Rapla &  24,108     &   0.755   & 0.570 & 0.463 \\
    Saare & 25,374      &   0.828    & 0.766 & 0.713 \\
    Tartu & 132,888     &   0.727    & 0.714 & 0.579 \\
    Valga & 17,528      &   0.772    & 0.635 & 0.549 \\
    Viljandi & 33,767   &   0.793    & 0.681 & 0.625 \\
    Võru & 28,405       &   0.777    & 0.710 & 0.668 \\
  \bottomrule                          
\end{tabular}
\caption{Statistics about the dataset. Users' counts are listed individually under four features (gender, age-groups, language, and location) to provide a comprehensive insight into the dataset. For example, under feature \textit{Languages}, users are categorized in three languages that is \textit{Estonian} (102,545 individual speakers), \textit{Russian} (14,882 individual speakers), and \textit{English} (236 individual speakers). Additionally, segregation values using \textit{FSI}, \textit{HI} and \textit{OBI} are reported on four features (gender, age-groups, language and location) using the CDR dataset which is explained in detail in Section \ref{sec:Evl}.}
\label{Table:DataSetStats}
\end{table}

\begin{figure*}
    \centering
    \includegraphics[width=1.5\columnwidth]{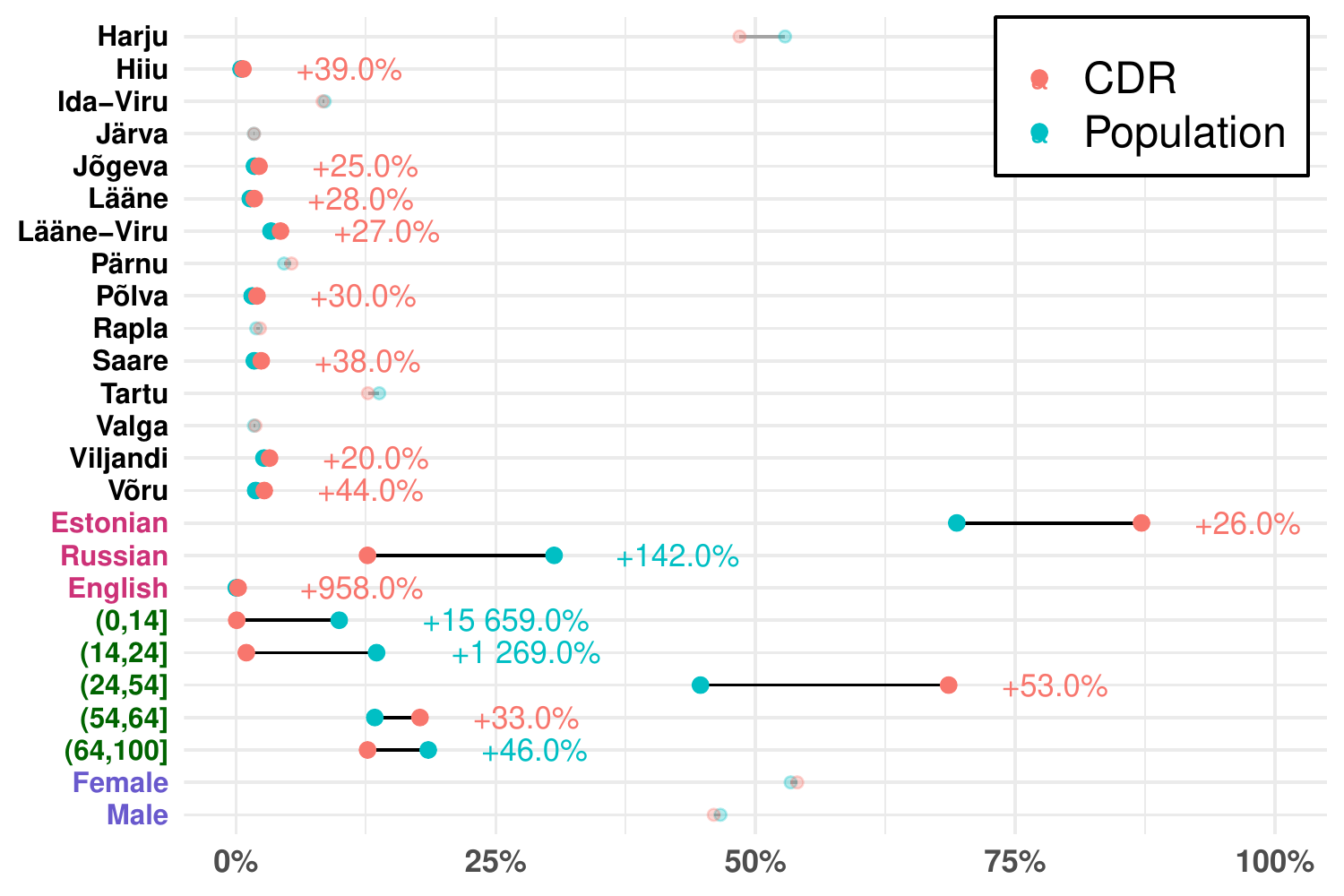}
    \caption{\textbf{Comparison of actual population of Estonia and users in \textit{CDR} based on four features.} Here, x-axis represents percentage and y-axis represents users' features from top to bottom namely \textit{location} (written in black), \textit{language} (highlighted in \textcolor{purple}{purple}), \textit{age-group} (highlighted with \textcolor[rgb]{0.0, 0.5, 0.0}{\textbf{green}}) and \textit{gender} (highlighted in \textcolor[rgb]{0.34, 0.17, 1.0}{blue}). The sum of all percentages based on each feature is 100 percent for both \textit{CDR} and actual Estonian population separately.  For each feature value, pink dots represent percentage of users in the \textit{CDR} dataset and seagreen dots represent percentage of people of actual Estonian population. Furthermore, for each feature value, the difference between \textit{CDR} and actual population percentage is calculated using formula \ref{eq:diffF}. All the differences greater than 20 percent are written and highlighted with text color. For example, for age-group \textcolor[rgb]{0.0, 0.5, 0.0}{(64,100]}, difference is \textcolor[cmyk]{0.7, 0, 0.20998, 0}{+46.0\%} which means actual population of Estonia is 46 percent more than the representation of users in \textit{CDR} dataset.}
    \label{fig:seg_dataset}
\end{figure*}

\textbf{Encoding users' age.} Centered on the official age-group categorization suggested by \textit{Europe-Bureau} and \textit{Statistics Estonia} \cite{estoniamean}, we categorize the age of users into the following five groups: (1) 0-14 years: Children; (2) 15-24 years: Early working age; (3) 25-54 years: Prime working age; (4) 55-64 years: Mature working age; and (5) 65+: Elderly.

In Table \ref{Table:DataSetStats}, the \textit{Age-Groups} row displays the distribution of users in the dataset according to their \textit{age-group}. E.g., for the age-group \textit{(24,54]}, 
there exist 83,028 individual users in the dataset that belongs to age-group (24,54].

\textbf{\textit{CDR} dataset validation using census dataset.} As \textit{CDR} data contains call records that might not represent the actual interaction among individuals, we need to validate that the \textit{CDR} data is indeed an acceptable representative of Estonia's actual population. We compare the distribution of users in the \textit{CDR} based on four features (\textit{location, language, age-group} and \textit{gender}) with the actual Estonian population in Figure \ref{fig:seg_dataset}, where, 
the x-axis represents percentage and the y-axis represents users' features. For each feature value (e.g., Harju, Estonian, (24,54], Female, etc.), the percentage of users in \textit{CDR} and actual population are plotted. Next, the difference between the \textit{CDR} percentage and actual population percentage is calculated using formula \ref{eq:diffF} for each feature value.
\begin{equation}\label{eq:diffF}
    Difference = \frac{|CDR\ \% - Population\ \%|}{min(CDR\ \%, Population\ \%)}*100
\end{equation}

The differences greater than 20 percent are written and highlighted with the text color. The features and the difference interpretation from top to bottom on the y-axis are as follows:

\noindent 1) \textit{\textbf{Location}}: There are 15 counties in Estonia, with the \textit{Harju} county, which contains the capital Tallinn being the most populated, followed by the \textit{Tartu} county as the second most populated, and the \textit{Hiiu} county as the least populous. From Figure \ref{fig:seg_dataset}, we can observe that the difference between \textit{CDR} users and the actual population in top-4 populated counties, representing nearly 80 percent of Estonia's real population
, is less than 20 percent\cite{estonia2012statistical}. All the counties with a difference of more than 20 percent cover approximately 14 percent of Estonia's population. Therefore, we can conclude that \textit{CDR} data is indeed a good representative of the actual Estonian population based on \textit{location}.
    
\noindent 2) \textit{\textbf{Language}}: As we mentioned earlier that the preferred language of interaction choices in the \textit{CDR} dataset 
are either \textit{Estonian}, \textit{Russian}, or \textit{English}. In Figure \ref{fig:seg_dataset}, we can observe that the representation of the Estonian-speaking population is higher in \textit{CDR} data than the actual Estonian population. On the other hand, this behaviour is reversed 
for the Russian-speaking population. The possible reason for such behavior is the fact that foreign-language speakers speak 
mostly 
with individuals from other countries and prefer to use other cheap mediums to connect, such as Whatsapp, Facebook, Skype, Telegram, etc\cite{taipale2018big}. Hence, we can conclude that \textit{CDR} data can be used 
for calculating segregation based on language.

\noindent 3) \textit{\textbf{Age-groups}}: Based on age-group, we observe that mobiles are mostly used by prime working age users (i.e., (24,54]), mature working age users (i.e., (54,64]), and elderly users (i.e., (64,100]). These three age-groups cover approx 99 percent of the \textit{CDR} dataset and 76 percent of the total Estonian population. Mobile usage percentages for other age-groups are relatively less than their actual population. From this, we can infer that in the \textit{CDR} dataset, the representation of the prime working age users is significant 
considering their actual population in Estonia. Thus, we can argue that the prime working age users' findings can be considered accurate with reasonable confidence. The same is valid for mature working age and elderly users. On the other hand, the representation of children and early working age-group in \textit{CDR} is less to negligible 
compared to the actual population, making it difficult to analyze these age-groups.

\noindent 4) \textit{\textbf{Gender}}: For both genders (\textit{female} and \textit{male}), the difference is less than 20 percent. Thus, we can infer that \textit{CDR} data is a good gender-based depiction of the real Estonian population.

    


Based on 
\textit{CDR} data validation using Estonian census data, we can infer that the \textit{CDR} is indeed a fair representative of Estonia's actual population, and all the study findings on the \textit{CDR} dataset can be deemed correct with reasonable certainty. 

\textbf{\textit{CDR} dataset availability. }Our dataset is partly location data, and it can not be shared due to privacy concerns. Additionally, despite the fact that the dataset is anonymized at two levels, there is still a small chance that specific persons can be identified. The dataset is owned by our university lab and is accessible for research purposes after signing the NDA. 

%% file: Methodology.tex
\section{Proposed Methodology}\label{sec:methodology}
This section first explains the motivation (in Section \ref{subsec:motivation}) behind proposing new indexes for measuring different aspect of segregation, \textit{Individual Segregation Index (\textbf{ISI})}, \textit{Individual Inclination Index (\textbf{III})}, and \textit{Overall Behavioural Index (\textbf{OBI})}. The \textit{OBI} is calculated 
using \textit{ISI}, and \textit{III}. The \textit{ISI} index measures whether segregation exists at individual level (Section \ref{subsec:ISI}), 
and \textit{III} index measures the inclination  (Section \ref{subsec:III}). 

\subsection{Motivation}\label{subsec:motivation}
As mentioned earlier in Section \ref{sec:intro}, the past indexes are over-simplified and use over-reduced information about groups to measure segregation; that is, they use highly aggregated information (the number of individuals in different groups and the ties within and in-between groups), which often results in an imprecise measurement of segregation. Also, many existing indexes (for example, the most well-known dissimilarity index \cite{duncan1955methodological}) does not look into the behaviour that results in segregation. In our work, we are using a separate index \textit{III} to measure the inclination that explains this behaviour by analyzing individuals' connection preferences.

Moreover, a few indexes measure both segregation and inclination simultaneously (e.g., homophily index \cite{coleman1958relational}. However, the information is over-reduced again, making it impossible to look at the behavior 
that resulted in the segregation. By behaviour, 
we mean that how individuals' of a group are connected to other groups. In this work, we propose the \textit{OBI} index, which measures the segregation (using \textit{ISI} and \textit{III} index), and it is possible to see the individuals' behaviour 
of the group of the population that results in the segregation. 
We can analyze the population's actual behaviour using \textit{OBI} index, which is not possible using other existing indexes.

\subsection{Individual Segregation Index (ISI)}\label{subsec:ISI}
We first describe annotations that will be used throughout this section. Let $G$ is a set of $n$ number of attribute groups i.e., $G$ = \{$g_1$, $g_2$, $g_3$ ... $g_n$\} and $A$ is a set of population distribution for each group \{$\alpha_1$, $\alpha_2$, $\alpha_3$ ... $\alpha_n$\}. Then, \textit{ISI} can be formalised as follows:
\begin{equation}\label{eq:ISI_method}
    ISI = \frac{\sum_{i=1}^{n} |P_{g_i} - \alpha_i|}{2*(1-min(A))}
\end{equation}

where, $P_{g_i}$ represents the proportion of an individual's connectivity to group $i$ and $min(A)$ is the minimum value in set $A$.

\begin{figure}[ht!]
    \centering
    \includegraphics[width=0.8\columnwidth]{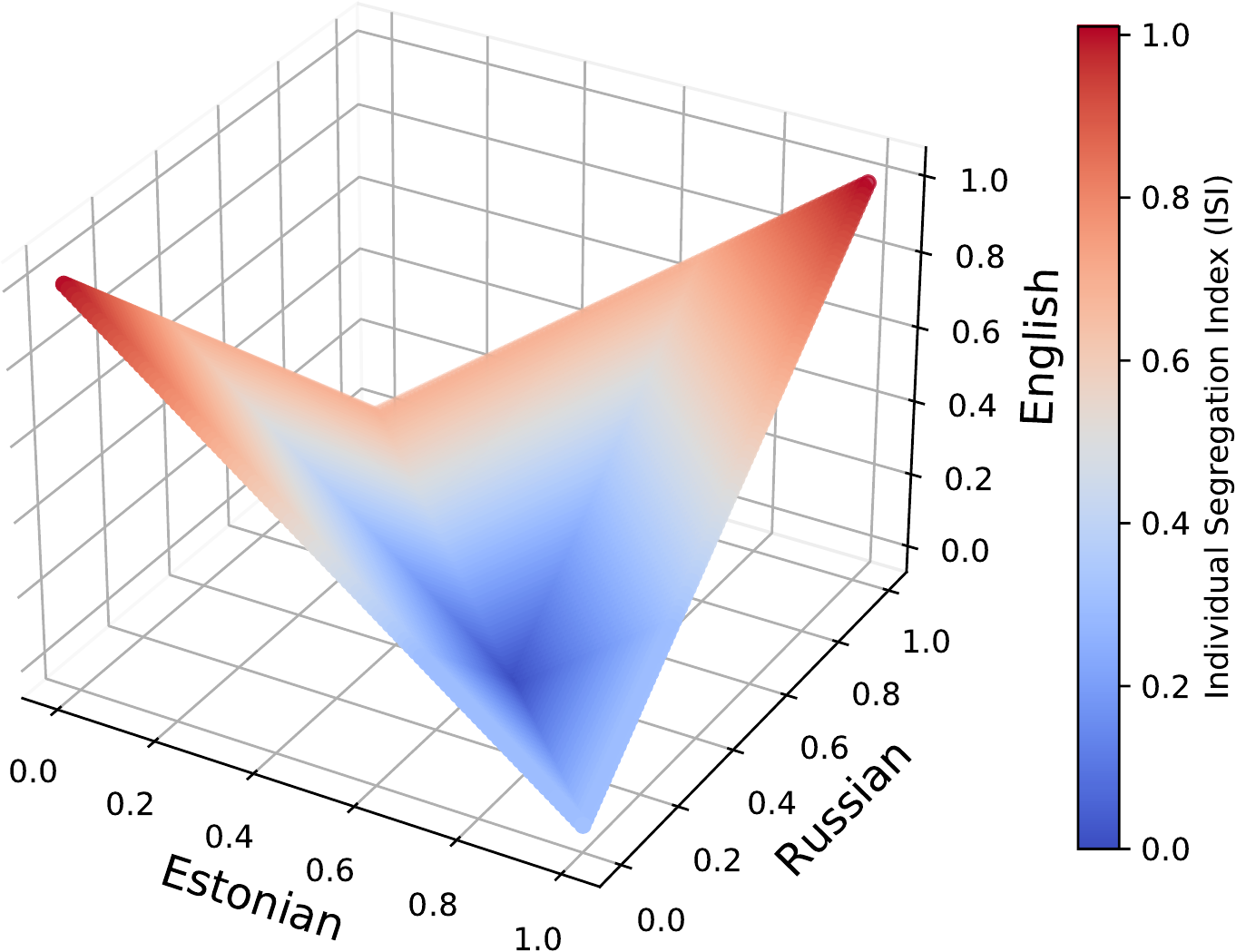}
    \caption{The \textit{ISI} index heatmap for an individual on different connection combinations to Estonian, Russian and English-speaking population in Estonia.}
    \label{fig:ISI_method}
\end{figure}

The \textit{ISI} index applies to any categorical variable (whether demographic or not), 
and its range varies between 0 and 1, where 0 means no-segregation (or individuals connectivity follows population distribution) and 1 means high segregation for the population group under investigation. For example, considering Estonia's three different language speaking population as groups, i.e., \textit{Estonian}, \textit{Russian} and \textit{English}-speaking groups. The population distribution of these groups in Estonia is 0.69, 0.3 and 0.01  \cite{estonia2012statistical}. Figure \ref{fig:ISI_method} shows the heatmap for \textit{ISI} index 
which clearly shows that an individual speaking to only English-speaking individuals is more segregated than an individual connected to only Russian-speaking individuals. Similarly, an individual speaking to only Russian-speaking individuals is more segregated than an individual connected to only Estonian-speaking individuals.

The \textit{ISI} index can calculate segregation; however, it fails to detect the segregation behavior's reasons. Next, we will define the index to identify the connection preference (or inclination) of the segregated individuals.

\subsection{Individual Inclination Index (III)}\label{subsec:III}
The \textit{III} index is useful to measure the connection preferences (or inclination) of an individual in a group of the population under investigation. Considering the notations defined in Section \ref{subsec:ISI}, the \textit{III} index for an individual that belongs to group $k$ can be defined as follows:

\begin{equation}\label{eq:III_method}
    III = \begin{cases} \frac{P_{g_k} - \alpha_k}{P_{g_k}} &: P_{g_k} - \alpha_k \leq 0 \\
    \frac{P_{g_k} - \alpha_k} {1-\alpha_k} &: P_{g_k} - \alpha_k \geq 0
    \end{cases}
\end{equation}

where, $P_{g_k}$ represents the proportion of an individual's connectivity to its own group $k$. Please note that for $P_{g_k} - \alpha_k = 0$, both use case 
are valid. 

The \textit{III} index is applicable to any categorical variable (whether demographic or not), 
and its range varies between -1 to 1, where -1 means complete inclination towards other groups and 1 means total inclination towards its own group. On the other hand, the value of 0 for \textit{III} index means no inclination towards any group. 
Please note that to calculate the \textit{III} index for an individual, we need to consider its group, which is not a requirement for calculating the \textit{ISI} index.

\subsection{Overall Behavioural Index (OBI)}\label{subsec:OBI}
The \textit{OBI} index 
can be used to understand the behaviour of an individual or a group of population. This is calculated using \textit{ISI} and \textit{III} index and can be formulated as follows:
\begin{equation}\label{eq:OBI_method}
    OBI = \begin{cases} sign(III)*\left(\frac{ISI + abs(III)}{2}\right) &: III \neq 0 \\
    (-1)*\left(\frac{ISI}{2}\right) &: III = 0
    \end{cases}
\end{equation}

where, $sign(III)$ and $abs(III)$ represents the sign and absolute value of the \textit{III} index respectively. For example, $sign(-0.4)$ will return $-1$ and $abs(-0.4)$ will give $0.4$.

\begin{figure}[ht!]
    \centering
    \includegraphics[width=0.9\columnwidth]{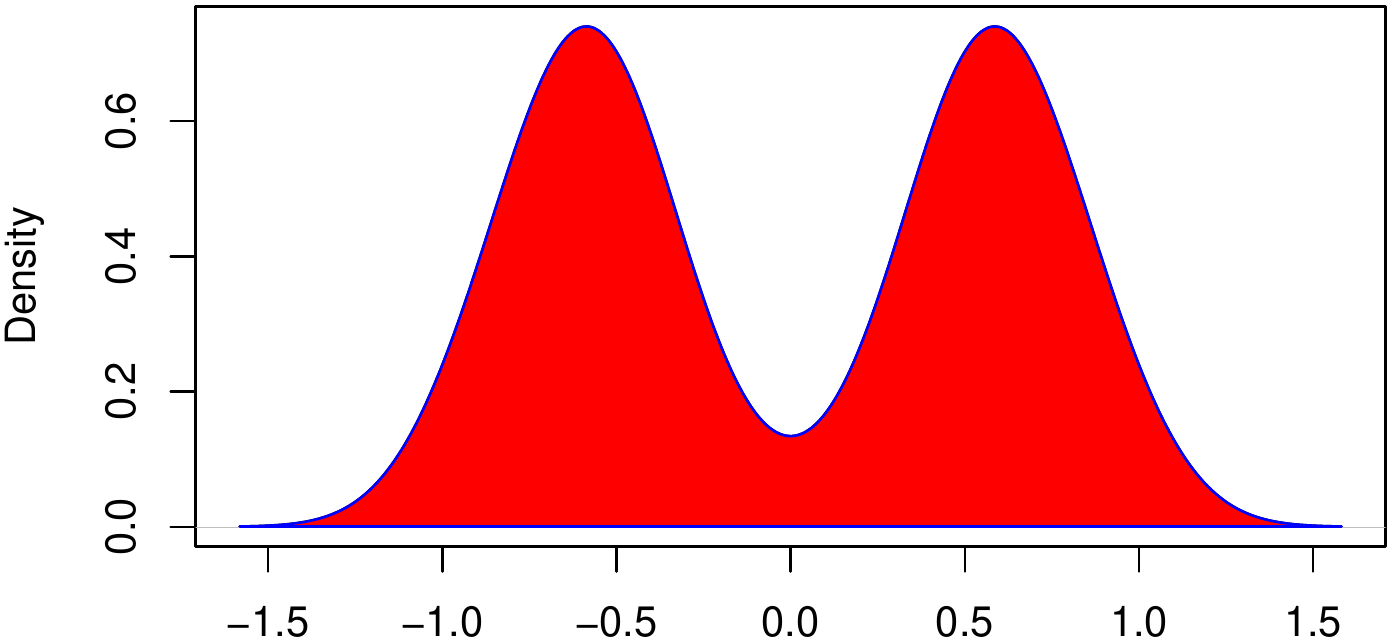}
    \caption{The \textit{OBI} index distribution for a group. The average of \textit{OBI} index for the group is zero.}
    \label{fig:III_method}
\end{figure}

The \textit{OBI} index considers the behavioural distribution of each individual in the group of the population, and its range varies between -1 to 1, where -1 means highly segregated with a complete inclination towards other groups and 1 also means highly segregated but with a complete inclination towards the same group. The value 0 for \textit{OBI} index means no segregation. Thus, the OBI index is non-simplified and gives more precise information on whether segregation exists or not. Please note that the average of \textit{OBI} index for a group can be zero in two possible ways. First, when individuals' behavioural distribution in a group follows actual population distribution. Second, if \textit{OBI} index for a group is symmetric about zero as shown in Figure \ref{fig:III_method}, that is \textit{OBI} and -\textit{OBI} have the same distribution, and can be written as \textit{OBI} $\overset{D}{=}$ -\textit{OBI}.

\subsection{Example}\label{subsec:ToyExample}
\begin{figure}
\minipage{0.45\columnwidth}
  \includegraphics[width=\linewidth]{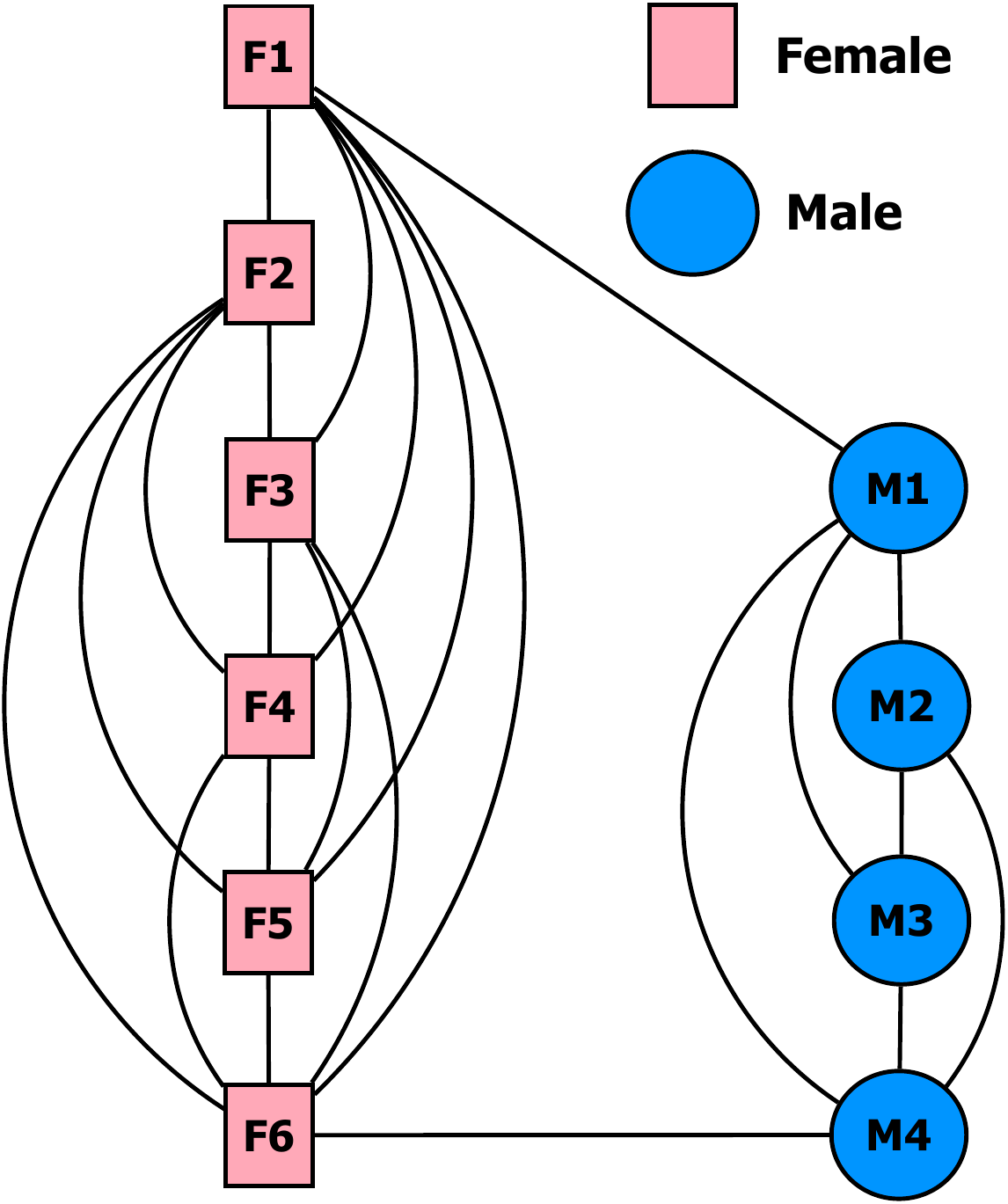}
  \caption{Connectivity network.}\label{fig:example2}
\endminipage\hfill
\minipage{0.48\columnwidth}
  \begin{tabular}{|c|c|c|c|}
\hline
{} & ISI & III & OBI \\\hline
F1 & 0.39 & 0.58 & 0.48 \\
F2 & 0.67 & 1 & 0.83 \\
F3 & 0.67 & 1 & 0.83 \\
F4 & 0.67 & 1 & 0.83 \\
F5 & 0.67 & 1 & 0.83 \\
F6 & 0.39 & 0.58 & 0.48 \\
M1 & 0.5 & 0.58 & 0.54 \\
M2 & 1 & 1 & 1 \\
M3 & 1 & 1 & 1 \\
M4 & 0.5 & 0.58 & 0.54 \\
\hline
\multicolumn{4}{c}{\textbf{Table 3: Index values.}}\\
\end{tabular}
\endminipage
\end{figure}

This section demonstrates how to calculate \textit{OBI} though a toy example consisting of two groups based on gender, i.e., \textit{female} and \textit{male} groups. The connectivity of each individuals is shown in Figure \ref{fig:example2}. As the population distribution among \textit{female} and \textit{male} groups are: 0.6, and 0.4 respectively, therefore, \textit{female} represents majority group and \textit{male} minority group.

\textbf{Calculating \textit{ISI} index. }Here, we calculate the \textit{ISI} index for 
female $F1$ in Figure \ref{fig:example2} that has connectivity to \textit{female}, and \textit{male} groups as 5/6 and 1/6 
respectively. The \textit{ISI} index for $F1$ can be calculated as: 
$$ISI_{F1} = \frac{|\frac{5}{6}-0.6| + |\frac{1}{6}-0.4|}{2*(1-min(0.6, 0.4))} = 0.39$$

Table 3 (Column 2) shows \textit{ISI} index for other females and males which clearly indicates that females ($F2$, $F3$, $F4$ and $F5$) are more segregated than females $F1$ and $F6$. However, males $M2$ and $M3$ are the most segregated individuals in the whole network because males $M2$ and $M3$ are only connected to other males, and \textit{male} group is a minority group. On the other hand, the females $F2$, $F3$, $F4$ and $F5$ are connected to only other females and are less segregated than males $M2$ and $M3$, because \textit{female} group is a majority group. Using \textit{ISI} index, we can identify individuals with high or low 
segregation; however, to understand their behaviour in-depth, we need to understand their inclination.

\textbf{Calculating \textit{III} index. }For calculating \textit{III} index, we need to consider the gender of the individual under investigation. For female $F1$ in Table 3 (Column 3), the \textit{III} index can be calculated using formula \ref{eq:III_method} as: $$III_{F1} = \frac{\frac{5}{6}-0.6}{1-0.6} = 0.58$$ 
Similarly, we can calculate \textit{III} index for other individuals as well. Here, we can observe that individuals only connected to the same group have the \textit{III} value as 1. 

\textbf{Calculating \textit{OBI} index. }We can calculate \textit{OBI} index for female $F1$ in Table 3 (Column 4) using formula \ref{eq:OBI_method} as: $$OBI_{F1} = sign(0.58)*\left(\frac{0.39 + abs(0.58)}{2}\right) = 0.48$$

\begin{figure}[ht!]
    \centering
    \includegraphics[width=0.8\linewidth]{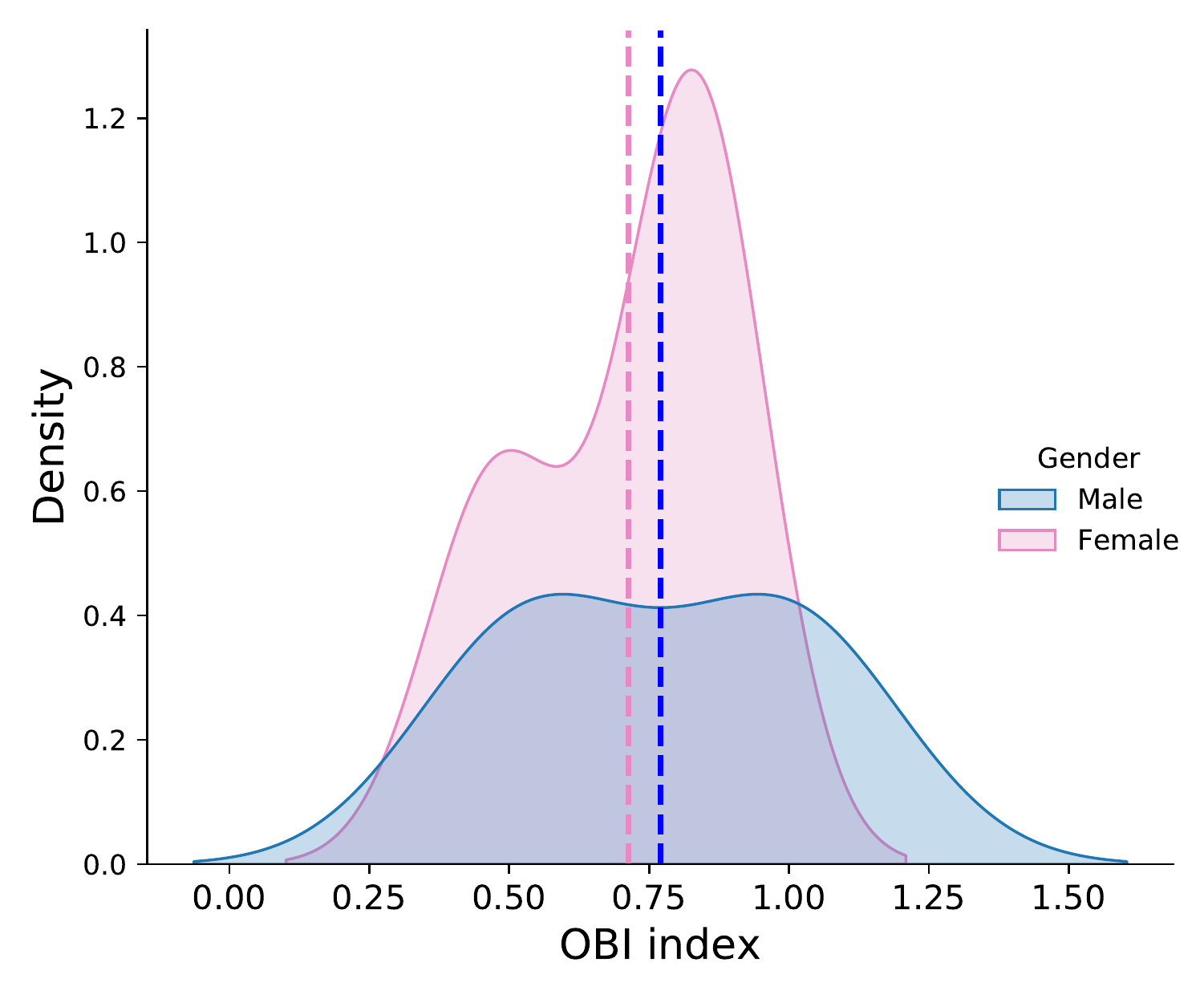}
    \caption{Distribution of OBI index for the Table 3.}\label{fig:OBIDistExample}
\end{figure}

Here, we can observe that considering the population distribution in Figure \ref{fig:example2}, a male only connected to \textit{male} group is the most segregated, and a female only connected to both \textit{female} and \textit{male} group is the least segregated. Figure \ref{fig:OBIDistExample} displays the \textit{OBI} index distribution for each individual, that highlights the behaviour of each group that results in segregation. The dotted lines represent the average of \textit{OBI} index for each group. 

%% file: Eval.tex
\vspace{-5pt}
\section{Evaluation}\label{sec:Evl}
\vspace{-5pt}
This section starts by briefly describing \textit{Freeman's segregation index} (Section \ref{subsub:FreemanIndex}) and \textit{homophily index} (Section \ref{subsub:CHI}), the two commonly used indexes for measuring segregation \cite{mele2020does}, which are used as baseline to compare with our proposed \textit{OBI} index. Next, in Section \ref{subsec:CDRAnalysis}, 
we measure and compare the \textit{OBI} with \textit{Freeman} and \textit{homophily index} 
using CDR data. 


\begin{figure*}[ht!]
\centering
\subfloat[Gender]{\includegraphics[width=0.35\linewidth]{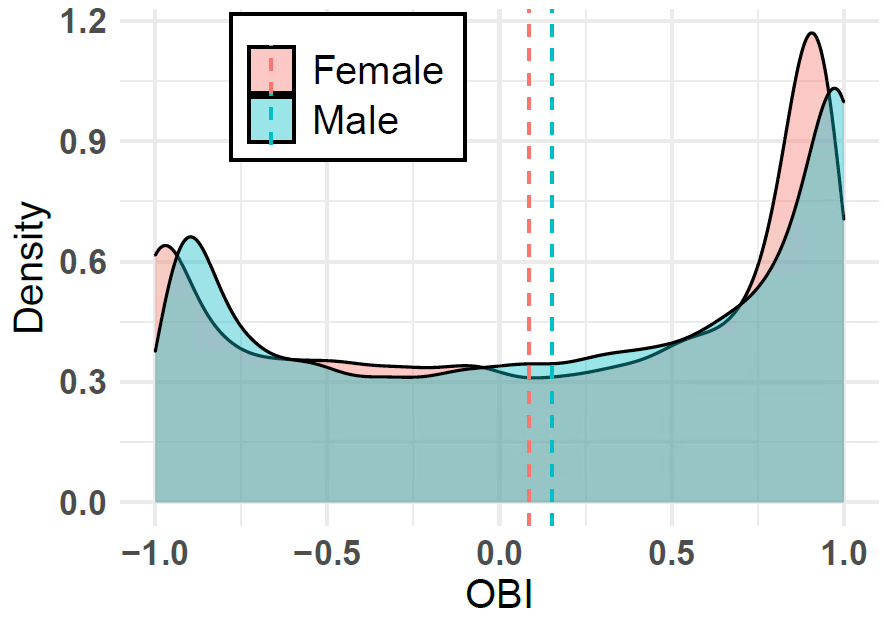}\label{fig:genderOBI}}
\hspace{3mm}
\subfloat[Age-Group]{\includegraphics[width=0.35\linewidth]{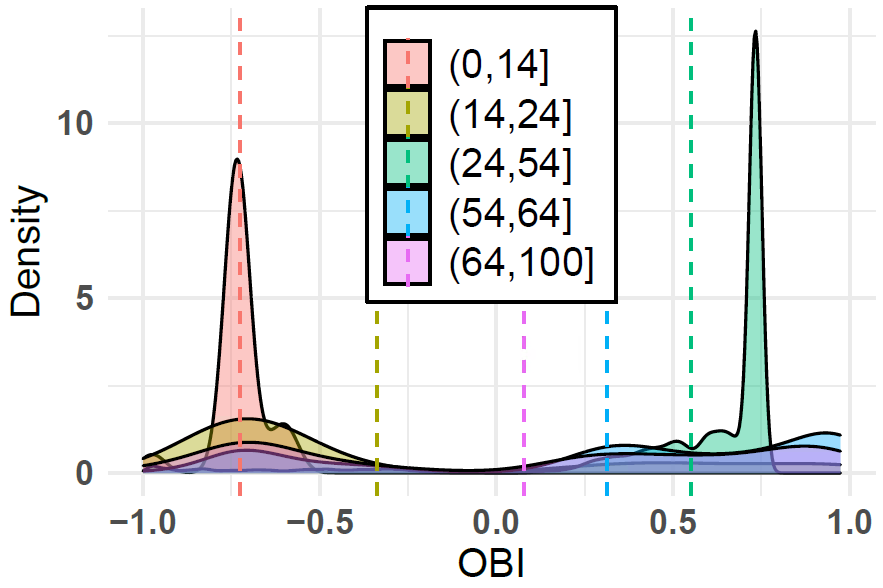}\label{fig:age_groupOBI}}\\
\subfloat[Language]{\includegraphics[width=0.35\linewidth]{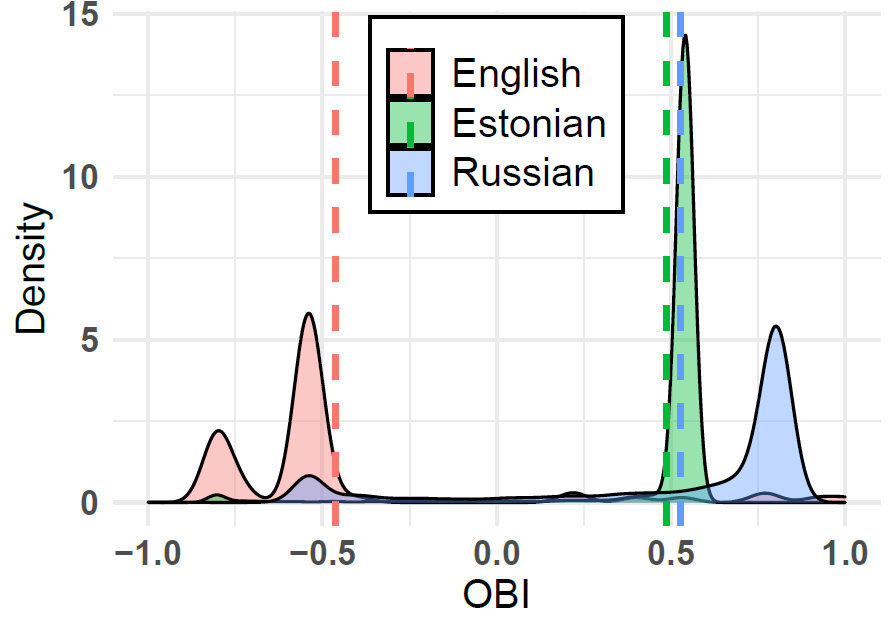}\label{fig:languageOBI}}
\hspace{3mm}
\subfloat[Location (County)]{\includegraphics[width=0.35\linewidth]{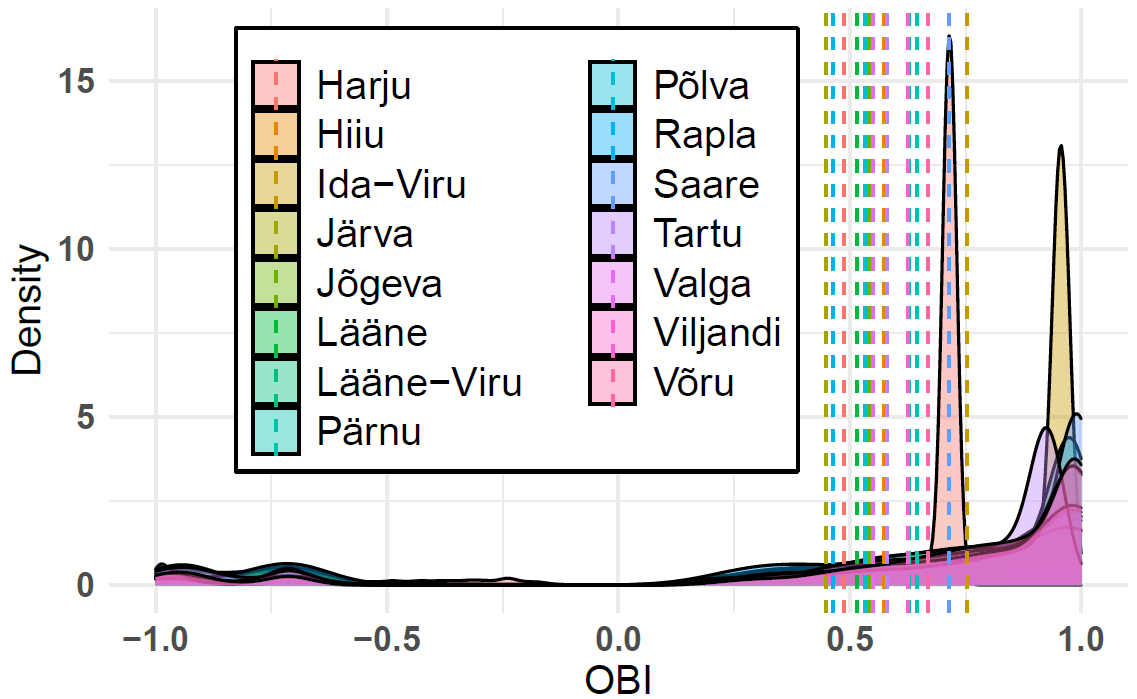}\label{fig:locationOBI}}
\caption{OBI on (a) gender, (b) age-groups, (c) language, and (d) location (county) features of the CDR dataset.}
\label{Fig:OBI_Eval}
\end{figure*}

\vspace{-3pt}
\subsection{\textbf{Freeman's Segregation Index}}\label{subsub:FreemanIndex}
\vspace{-5pt}
The Freeman Segregation Index (FSI) has been extensively used for understanding segregation in the social interaction network \cite{freeman1978segregation}. According to FSI, if a given attribute (group label) does not apply to social connection 
then connections should be randomly distributed with respect to the attribute. Thus, the disparity between 
cross-group ties expected by chance and 
observed is used to measure segregation. 

\textbf{Calculating FSI value.} Let us consider a network with static attribute groups \textit{A} and \textit{B} (of relative size $N_A$ and $N_B$ with $N_A + N_B=1$) distributed among nodes uniformly at random and independently of the network structure, such that there is a fraction $P_{AB}=P_{BA}$ of edges between groups, and fractions $P_{AA}$, $P_{BB}$ within each group ($P_{AA}+P_{AB}+P_{BB}=1$). The \textit{FSI} can be measure using below formula: 
\begin{equation}\label{eq:FSI}
    FSI = 1-\frac{X}{E(X)}
\end{equation}

where, \textit{X} is the proportion of between group ties and \textit{E(X)} is the expected proportion of random ties. The \textit{X} and \textit{E(X)} can be calculated using formula \ref{eq:X} and \ref{eq:EX} respectively.
\begin{equation}\label{eq:X}
    X = \frac{P_{AB}}{P_{AA}+P_{AB}+P_{BB}}
\end{equation}
\begin{equation}\label{eq:EX}
    E(X) = \frac{2 N_A N_B}{(N_A + N_B)(N_A + N_B - 1)}
\end{equation}

\subsection{\textbf{Homophily Index For Measuring The Segregation}}\label{subsub:CHI}
Homophily is the tendency of individuals to interact and associate with other individuals. 
In the past, homophily has been studied in great detail in numerous works \cite{asikainen2020cumulative}. 
These studies indicate that the similarity is correlated with the connection among individuals and can be categorized based on age, 
gender, 
class, 
ethnicity \cite{sahasranaman2018ethnicity}, etc. In this work, we use the Coleman homophily index (\textit{HI}) \cite{coleman1958relational} for comparison with OBI since \textit{HI} is commonly used to compares the homophily of groups with different sizes by normalizing the excess homophily of groups by its maximal value \cite{coleman1958relational}.

\noindent \textbf{Calculating \textit{HI} value: }Considering the notations defined in Section \ref{subsub:FreemanIndex}. In the case of two attribute groups, the probability that a random edge from a node in a group \textit{A} leads to a node in group \textit{A} is defined as:
\begin{equation}
    T_{AA} = \frac{2P_{AA}}{2P_{AA} + P_{AB}}
\end{equation}

Similarly, we can write equation for $T_{BB}$. The \textit{HI} value for group $A$ ($HI_A$) and $B$ ($HI_B$) can be calculated using 
\vspace{-5mm}

\begin{multicols}{2}
  \begin{equation}
        HI_{A} = \frac{T_{AA} - N_A}{1 - N_A}
    \end{equation}\break
    \begin{equation}
        HI_{B} = \frac{T_{BB} - N_B}{1 - N_B}
    \end{equation}
\end{multicols}

The range for both $HI_A$ and $HI_B$ is from -1 to 1, where -1 for $HI_A$ means that group $A$ individuals only connects with group $B$ individuals (only in between groups connections), whereas 1 for $HI_A$ means that group $A$ individuals only connects with group $A$ individuals (only within-group).

\subsection{Measuring Segregation Using CDR Dataset}\label{subsec:CDRAnalysis}
\vspace{-2mm}
The central goal of segregation indexes is to measure the degree of separation between two or more population groups. 
When measuring segregation using sampled segregation indexes, one would expect that the relative order of segregation is preserved. For example, in Table \ref{Table:DataSetStats} (for feature \textit{Languages}), the order of segregation according to all segregation indexes (\textit{FSI}, \textit{HI} and \textit{OBI}) is \textit{Russian}$<$\textit{Estonian}$<$\textit{English}, that is, the Russian-speaking population is the most segregated and the English-speaking population is the least segregated. This property of preserving relative order of segregation can be called \textit{consistency} and formalized as follows:

\textbf{(Consistency definition)}. \textit{Let the connection network data D be fixed. Any number of segregation indexes $I_1$, $I_2$, $I_3$ ... $I_n$ are consistent on any features (say X and Y), if the relative order of all indexes are preserved. That is,}

\begin{equation}
    I_k(D,X)>I_k(D,Y); \hspace{3mm} \forall_{k\in\{1,2,..,n\}}
\end{equation}

where, $I_k(D,X)$ is the segregation value using segregation index $I_k$ on data $D$ and feature $X$.

Consider one more example, in Table \ref{Table:DataSetStats} (for feature \textit{Age-Groups}), age group (64,100] is most segregated according to \textit{FSI}, on the other hand, both \textit{HI} and \textit{OBI} find out that age-group (24,54] is the most segregated. Here, \textit{consistency} is not reserved. Thus analyzing individual's behaviour is necessary to report accurate segregation that is possible using \textit{OBI}.

Figure \ref{Fig:OBI_Eval} visualize the individuals' OBI index distribution based on gender (\ref{fig:genderOBI}), age-group (\ref{fig:age_groupOBI}), language (\ref{fig:languageOBI}) and location (\ref{fig:locationOBI}). For example, \textit{OBI index} distribution on \textit{Age-Groups} shows that children (i.e., (0,14]) and early-working age (i.e., (14,24]) population are more connected to other age groups (see Figure \ref{fig:age_groupOBI}). On the other hand, prime-working age (i.e., (24,54]), mature-working age (i.e., (54,64]) and elderly (i.e., (64,100]) are more inclined towards same age group individuals. On comparing the \textit{OBI index} on \textit{Age-Groups} with \textit{HI index} (see Table \ref{Table:DataSetStats}), we observe that \textit{HI index} is not able to measure accurate segregation for children (i.e., (0,14]) and early-working age (i.e., (14,24]) population. Hence, 
this shows that \textit{OBI index} can measure segregation more precisely than the other baseline indexes. 
\vspace{-2mm}

%% file: Conclusion.tex
\section{Discussion and Conclusion}\label{sec:concl}
\vspace{-1mm}
Segregation indexes help in summarizing the relationship between various groups and provide a basis for public policy intervention. But, finding an ideal segregation index from the existing set of indices is challenging for the following reasons. First, due to the complexity of varied dimensions and arrangements in society. 
Second, due to the \textbf{over-simplified and over-reduced} nature of the past measures. To overcome this, we proposed 
\textbf{O}verall \textbf{B}ehavioural \textbf{I}ndex (OBI) 
segregation index which is 
combination of \textit{Individual Segregation Index (ISI)} and \textit{Individual Inclination Index (III)} that reports the segregation as well as the inclination of the population group under investigation. 
Thus, the \textit{OBI index} is non-simplified and makes it possible to analyze the connectivity behaviour of individuals of groups that leads to segregation.

We validated the CDR dataset with Estonian census data 
while interpreting the results and reporting the segregation. To further maintain the data quality, we do not consider the individuals with fewer connections during the analysis. We applied filters at two stages. First, we used the statistical test to identify the significant number of connections, and 
based on which the individuals with less than six connections are not considered. Second, the individuals with at least six connections with other individuals with known features are considered. 
A possible limitation of 
the \textit{OBI index} is that, it is
relatively time-consuming (compared to past indices), as we are avoiding the over-simplification and over-reduction. 
Thus, reducing the time-complexity of the proposed method is an interesting direction for future research.